\documentstyle[graphicx,axodraw,12pt]{article} 
\begin{document}

\baselineskip 24pt

\newcommand{\sheptitle}
{Brane Mediated Supersymmetry Breaking}

\newcommand{\shepauthor}
{S. F. King
and D. A. J. Rayner}

\newcommand{\shepaddress}
{Department of Physics and Astronomy,
University of Southampton, Southampton, SO17 1BJ, U.K.}

\newcommand{\shepabstract}
{We propose a mechanism for mediating supersymmetry breaking 
in Type I string constructions. The basic set-up consists of 
a system of three D-branes: two parallel D-branes, 
a matter D-brane and a source D-brane, with
supersymmetry breaking communicated via a third D-brane,
the mediating D-brane, which
intersects both of the parallel D-branes.
We discuss an example in which the first and second family matter fields
correspond to open strings living on the intersection
of the matter D-brane and mediating D-brane, while
the gauge fields, Higgs doublets and third family matter
fields correspond to open strings living on the mediating 
D-brane. As in gaugino mediated models,
the gauginos and Higgs doublets receive direct soft masses
from the source brane, and 
flavour-changing neutral currents are naturally suppressed
since the first and second family squarks and sleptons 
receive suppressed soft masses.
However, unlike the gaugino mediated model, the
third family squarks and sleptons
receive unsuppressed soft masses, resulting in a
very distinctive spectrum with heavier
stops, sbottoms and staus. }

\begin{titlepage}
\begin{flushright}
hep-ph/0012076\\
\end{flushright}
\vspace{0.5in}
\begin{center}
{\Large{\bf \sheptitle}}
\vspace{0.5in}
\bigskip \\ \shepauthor \\ \mbox{} \\ {\it \shepaddress} \\
\vspace{0.5in}
{\bf Abstract} \bigskip \end{center} \setcounter{page}{0}
\shepabstract
\end{titlepage}


\section{Introduction} 

The process of SUSY breaking continues to be an active area of
research.  Over the years there have been various mechanisms proposed,
including gauge \cite{gauge} and
anomaly mediation \cite{anomaly}.  An alternative mechanism has been
been put forward \cite{kaplan,chacko} called {\it gaugino} mediated 
supersymmetry (SUSY) breaking which 
has the attractive property of solving the flavour problem, since
scalars masses effectively vanish at the GUT scale and 
are generated through radiative corrections for which a GIM-like mechanism
prevents flavour-changing neutral current
(FCNC) problems. 
This is rather like the no-scale supergravity 
mechanism \cite{no-scale}, 
but is implemented within a Horava-Witten \cite{witten}
type set-up
\footnote{Note that in the Horava-Witten model gauge fields
do not live in the bulk.}
consisting of two parallel but spatially separated D3-branes
with SUSY broken on one brane, with the SUSY matter fields living
on the other brane and the gauge sector living in the bulk and
communicating the SUSY breaking from one brane to the other.
The Higgs doublets may also be in the bulk providing a
solution to the $\mu$ problem via the Giudice-Masiero mechanism
\cite{giudicemasiero}.
The advantage of this set-up is that the contact terms arising from
integrating out states with mass $M$ are suppressed by a Yukawa
factor $e^{-Mr}$ if $M \ge r$, and so a modest separation between the 
two branes can lead to negligible direct communication between
the SUSY breaking brane and the matter brane. This is the starting
point of both the anomaly mediated and the gaugino mediated
models, and underpins the solution to the FCNC problem in both
cases.

In this paper we shall propose a mechanism for 
mediating SUSY breaking in Type I string models based on 
open strings starting and ending on D-branes. 
Type I string theories can provide an attractive setting
for ideas such as gaugino mediated SUSY breaking ($\tilde{g}MSB$),
and we shall explore this possibility in this paper.
In place of the Horava-Witten set-up we shall consider a Type I 
toy model consisting of 
two parallel D-branes with a third D-brane intersecting with both
of the parallel D-branes. Instead of having the gauge fields in the
bulk we shall put the gauge fields onto the third mediating
D-brane, which allows SUSY breaking to be communicated between the
SUSY breaking brane and the matter brane.
Thus the role of the bulk is played by the third mediating D-brane,
and it is the gauge fields which live on this brane 
that communicates the SUSY breaking. However in Type I models
it is natural for a matter family to also live on the mediating
D-brane, and this provides a characteristic signature of the 
brane mediated SUSY breaking mechanism.

To illustrate these ideas we consider a toy model inspired by
the work of Shiu and Tye \cite{shiutye} using 
intersecting D5-branes, where the intersection regions are effectively
parallel D3-branes within a higher-dimensional spacetime. In this model
two chiral families occur in the 4d intersection region 
at the origin fixed point ($5_{1}
5_{2}$ sector), with a third family on the $D5_{2}$-brane ($5_{2}5_{2}$
sector). However our model differs from Shiu-Tye since we include 
a further $D5_1'$-brane which intersects
with the $D5_2$-brane at a point located
away from the origin fixed point, and suppose that SUSY gets broken 
on that brane and is communicated via the states
on the $D5_{2}$-brane which intersect with both $D5_1$-branes
at the two fixed points -- brane mediated SUSY breaking (BMSB).
In this example gauge fields, Higgs fields and the third family
all live in the mediating D-brane which plays the role of the bulk
in the original scenario. This separation of the third
family\footnote{Remember that the first two families are localised
within an effective 4d overlapping region, while the third family
feels two extra dimensions} 
provides an explanation for the large mass
of the third family of quarks and leptons,
without perturbing the solution to the flavour problem since the first and
second families remain almost degenerate.

The layout of the remainder of the paper is as follows.  In section 
\ref{sec:gauginomed} we review $\tilde{g}MSB$, and in section \ref{sec:typei}
we introduce a Type I string-inspired toy model motivated by Shiu and
Tye.  Section \ref{sec:bmsb} is the main section of the paper in which we 
present our toy model that illustrates the BMSB mechanism,
and explore its theoretical and experimental consequences.
Section \ref{sec:conc} concludes the paper.

\section{Gaugino Mediated Supersymmetry Breaking}  
\label{sec:gauginomed}

\begin{figure}[h]   
 \begin{center}  
  \begin{picture}(420,190)(0,0)
   \Line( 100, 110 )( 145, 185 )
   \Line( 100,  15 )( 145,  90 )
   \Line( 100,  15 )( 100, 110 )
   \Line( 145,  90 )( 145, 185 )
   \Text( 100,   5 )[c]{$x_5=0$}
   \Text(  90, 110 )[r]{``matter D3-brane''}
   \Text(  90,  90 )[r]{MSSM matter fields}
   \DashLine( 137, 155 )( 130, 114 ){4} 
   \CArc( 190, 10 )( 120, 60, 120 ) 
   \PhotonArc( 190, 10 )( 120, 60, 120 ){3}{10} 
   \Line( 130, 114 )( 120, 77 ) 
   \DashLine( 108, 45 )( 120, 77 ){4} 
   \CArc( 180, 180 )( 120, 240, 300 ) 
   \PhotonArc( 180, 180 )( 120, 240, 300 ){3}{10}
   \CArc( 238, 97 )( 21, 278, 412 ) 
   \PhotonArc( 238, 97 )( 21, 278, 412 ){3}{5} 
   \put( 251, 114 ){\circle*{6}} \put( 240, 76 ){\circle*{6}} 
   \Line( 225, 110 )( 270, 185 ) 
   \Line( 225, 15 )( 270, 90 )
   \Line( 225, 15 )( 225, 110 ) 
   \Line( 270, 90 )( 270, 185 )
   \Text( 225, 5 )[c]{$x_5=L$} 
   \Text( 280, 110 )[l]{``source D3-brane''} 
   \Text( 280, 90 )[l]{SUSY breaking sector} 
  \end{picture}
 \end{center} 
  \caption{{\small An extra dimensional loop diagram that contributes to
   SUSY breaking scalar masses.  It is similar to a self-energy diagram,
   but with the virtual gaugino not confined to either 4-dimensional brane.
   This Figure is taken from Ref. \cite{kaplan}.}} 
   \label{fig:loop}
\end{figure}
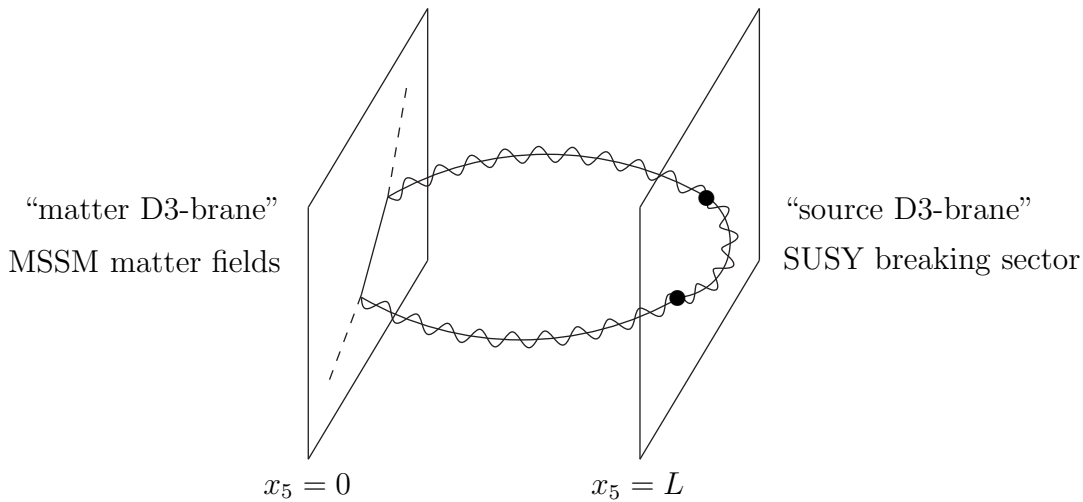

In this section we review the $\tilde{g}MSB$ mechanism in Refs. 
\cite{kaplan,chacko}.
This toy model involves D3-branes embedded in a higher-dimensional space.
Two parallel D3-branes are spatially separated along (at least) one extra 
dimension as shown in Fig. \ref{fig:loop}.  Standard Model quark and lepton 
fields are localised on the matter brane as open strings, while the gauge and
(possibly) Higgs fields propagate in the bulk\footnote{Thus feeling
all 5-dimensions.}.  Supersymmetry is broken on the displaced
source D3-brane.  SUSY breaking is communicated to the bulk fields by
direct higher-dimensional interactions\footnote{Higher-dimensional
operators are assumed to arise from the underlying string theory,
although this is not clear at present.}, and mediated to the
quark/lepton fields by Standard Model loops\footnote{Gauginos in the
bulk couple directly to chiral fermions on the matter brane.  They also
couple to the hidden sector directly through mass-insertions on the 
source brane.}.

The full D-dimensional lagrangian is split into two distinct pieces
- a bulk term involving only bulk fields and terms localised on either
D3-brane that allow direct bulk-brane field coupling.

\begin{equation}
  {\mathcal L}_{D} = {\mathcal L}_{bulk} \left( \Phi_{bulk} 
   \left( x,y \right) \right) + \sum_{j} \delta^{D-4}
   \left( y-y_{j} \right) {\mathcal L}_{j} \left( \Phi_{bulk} \left(
   x,y_{j} \right), \phi_{j} \left( x \right) \right)
\end{equation}
   where j runs over the branes, x are coordinates for the 4
   non-compact dimensions, y are coordinates for the $D-4$ compact
   spatial dimensions, $\Phi_{bulk}$ is a bulk field, and $\phi_{j}$
   is a field localised on the $j^{th}$ brane.
 
A Naive Dimensional Analysis (NDA) allows the 5d (or higher)
effective theory to be {\it matched} on to the observed 4d theory at
the compactification scale.  The 4 and D-dimensional gauge couplings
can be related by the size of the compact dimensions:
\begin{equation}  \label{eq:gscaling}
   g_{4}^{2} = \frac{ g_{D}^{2} }{V_{D-4}}  
\end{equation}

The D-dimensional gauge coupling $g_{D}$ must be smaller than its
strong-coupling limit, otherwise perturbative results become
meaningless\footnote{Extra dimensions (and Kaluza-Klein excitations)
change the energy-dependence of couplings to power law running above
the compactification scale.  This allows for unification at lower
scales, see \cite{dienes} for a review.}
\begin{equation}  \label{eq:gdscaling}
  g_{D}^{2} \sim \frac{ \epsilon l_{D} }{ M^{D-4} }
\end{equation}
where $l_{D}$ is a geometrical loop factor for D dimensions, $l_{D} =
2^{D} \pi^{D/2} \Gamma(D/2)$, M is the fundamental scale in the theory
which acts as a regulating cutoff, and $\epsilon$ suppresses the
coupling strength.  $\epsilon \sim 1$ corresponds to the strong
coupling limit.  This places a constraint, along with FCNC
suppression, that restricts the maximum size of the extra dimensions.
(See \cite{kaplan,chacko} for details.)

Following the work of Randall and Sundrum on spatially-separated
D3-branes in extra dimensions \cite{anomaly}, contact terms between
fields on opposite branes are exponentially suppressed by an amount
$e^{-ML}$, where L is the separation between D3-branes along the extra
dimension(s).

Eq. (\ref{eq:contact}) is an example of an exponentially suppressed 4-point operator involving 
fields from the matter and source branes that generates scalar masses:
\begin{equation} \label{eq:contact} 
  \triangle {\mathcal L}_{brane} \sim \frac{ e^{-ML} }{ M^{2} } \int d^{4}
   \theta \left( \hat{\phi}^{\dagger}_{S} \hat{\phi}_{S} \right) 
   \left( \phi^{\dagger}_{M} \phi_{M} \right)
\end{equation}
\begin{center}
(where $\phi_{S}, \phi_{M}$ are source and matter fields respectively)
\end{center}
Compare the suppressed contact terms with the operators giving rise to
gaugino masses and Higgs SUSY breaking parameters, from Higgs
superfields $h_{u}, h_{d}$ and gauge field strengths $W_{\alpha}$
living in the bulk\footnote{The scale factors M arise from the
requirement of canonical normalization.}.

\begin{eqnarray} \label{eq:branel1} 
  \triangle {\mathcal L}_{brane} \sim \frac{ l_{D} }{ l_{4} } 
   \left( \int d^{2} \theta \frac{1}{ M^{D-3} } \hat{\phi}_{S} W^{\alpha}
   W_{\alpha} + h.c. \right) + \frac{ l_{D} }{ l_{4} } \int d^{4} \theta 
   \left\{ \frac{1}{ M^{D-3} } \left( \hat{\phi}^{\dagger}_{S} h_{u} h_{d} 
   + h.c. \right) \right. \nonumber \\ 
  + \left. \frac{1}{ M^{D-2} } \hat{\phi}^{\dagger}_{S} \hat{\phi}_{S}
   \left[ h_{u}^{\dagger} h_{u} + h_{d}^{\dagger} h_{d} + 
   \left( h_{u} h_{d} + h.c. \right)
   \right] \right\} 
\end{eqnarray}

This leads to soft terms when we match to the D-dimensional
theory\footnote{Notice that the $B\mu$ term and Higgs mass-squared terms 
are enhanced by a volume factor relative to the $m_{\lambda} , \mu$
terms.} and using eqs. (\ref{eq:gscaling},\ref{eq:gdscaling})
with $g_{4} \approx 1$:

\begin{equation} \label{eq:branel2}
   m_{\lambda} , \mu \sim \frac{ \hat{F}_{S} }{ M } 
    \frac{ l_{D}/l_{4} }{ M^{D-4}V_{D-4} } \sim \frac{1}{\epsilon l_{4}} 
    \frac{ \hat{F}_{S} }{ M } , \hspace{1cm}
   B\mu , m_{h_{u}}^{2} , m_{h_{d}}^{2} \sim \frac{ \hat{F}_{S}^{2} }{ M^{2} }
    \frac{ l_{D}/l_{4} }{ M^{D-4}V_{D-4} } \sim \frac{1}{\epsilon l_{4}} 
    \frac{ \hat{F}_{S}^{2} }{ M^{2} }
\end{equation}

Both papers discuss methods of generating the $\mu$-term\footnote{Ref. \cite{kaplan}
has the Higgs fields localised on the matter D3-brane, while \cite{chacko} has the
Higgs fields living in the bulk.}.  Ref. \cite{kaplan} suggested the 
inclusion of an additional gauge singlet on the matter brane (NMSSM) with an
extra superpotential term ${\mbox W \sim \lambda N h_{u} h_{d}}$.  An
effective $\mu$-term is produced if N acquires a non-zero vacuum expectation 
value (vev).
Another possibility \cite{chacko} is to produce the $\mu$-term on the source 
brane through the Giudice-Masiero mechanism \cite{giudicemasiero} (as above) 
${\mathcal L} \sim \int d^{4} \theta \lambda_{\mu} \hat{\phi}^{\dagger}_{S} 
h_{u} h_{d}$.

\section{Type I String-Inspired Model} \label{sec:typei}

Now we turn to Type I string constructions and introduce a toy model
motivated by the work of Shiu and Tye \cite{shiutye}.
The string scale $m_{s}$ is usually considered to be of the order
$10^{16}$ GeV, but recently the gauge unification scale was suggested
to be as low as 1 TeV, which could allow the string scale at a
comparable value.  Shiu and Tye \cite{shiutye} discuss the
phenomenological possibilities within Type I string theory and
overlapping D5-branes.  They use the duality between the
compactification of 10-dimensional Type IIB string theory on an
orientifold, with Type I theory on an orbifold to recover a
4-dimensional ${\mathcal N}=1$ supersymmetric chiral string model with
Pati-Salam-like gauge symmetry.

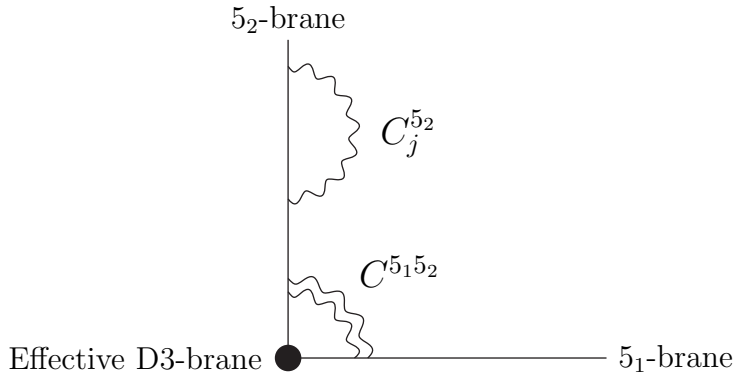
\begin{figure}[h]
 \begin{center}
  \begin{picture}(150,130)(5,5)
     \Line(5,5)(5,125)
     \Line(5,5)(125,5)
     \Vertex(5,5){5}
     \Text(-5,5)[r]{Effective D3-brane} 
     \Text(5,130)[b]{$5_{2}$-brane}
     \Text(130,5)[l]{$5_{1}$-brane}
     \PhotonArc(5,5)(25,0,90){2}{4}
     \PhotonArc(5,5)(30,0,90){2}{4}
     \Text(40,90)[l]{{\large $C^{5_{2}}_{j}$}}
     \Text(32,32)[lb]{{\large $C^{5_{1}5_{2}}$}}
     \PhotonArc(5,90)(25,270,90){2}{8} 
   \end{picture}
  \end{center}
   \caption{{\small The matter fields and Higgs doublets resulting
       from Shiu and Tye's third scenario with intersecting D5-branes, where 
       $ C_{i}^{p} $ is an open-string state (matter field) starting and 
       ending on the $p^{th}$ brane. $ C^{p q} $ is an open-string state 
       starting on the $p^{th}$ brane and ending on the $q^{th}$ brane.}}
\label{fig:2}
\end{figure}

Tadpole cancellations and a non-zero background NS-NS B-field
constrain the number and type of D-branes allowed within the model to
D5 and D9-branes only \cite{kaku}. In a particular scenario they
consider only one type of D5 brane $(5_3)$ together with the
D9 brane, and after T-dualizing they arrive at a scenario
with two intersecting branes, namely $5_1$ and $5_2$ branes
which intersect at the origin fixed point.
A gauge group $U(4) \otimes U(2)
\otimes U(2)'$ exists on each brane, and they discuss three
scenarios where the Standard Model gauge group originates from
different brane combinations.  Their third scenario is of particular
interest since it leads to three chiral families - two families on the
$5_{1} 5_{2}$ overlap and a third family on the $D5_{2}$-brane as shown in
Fig. \ref{fig:2}.

We can express the allowed superpotential \cite{ibanez} in terms of the
possible states from the two types of D5-branes present in this model:

\begin{equation} \label{eq:superp2} 
    W = C_{1}^{5_{2}} C_{2}^{5_{2}} C_{3}^{5_{2}} +
        C_{3}^{5_{2}} C^{5_{1} 5_{2}} C^{5_{1} 5_{2}}
\end{equation}

We now proceed to introduce a toy model based on the above
construction. \footnote{ For other examples of toy models based
on this construction see \cite{gordy} and references therein.}
In order to allow the third family Yukawa couplings 
$\left( \bar{F}_{3} F_{3} h \right)$ consistent with the string
selection-rules in eq. (\ref{eq:superp2}), we shall assign the Higgs
$ h_{u}, h_{d} \equiv C_{1}^{5_{2}}$ or $C_{2}^{5_{2}}$.  This leads to
the four possible allocations of $5_{2}$ states in Table \ref{tab:states1}.

\begin{table}[h]  
 \begin{center} 
  \begin{tabular}{|c||c|c|c|c|} \hline
   $5_{2}$ states & A & B & C & D \\ \hline \hline
     \vrule width 0pt height 13pt
   $h \sim h_{u},h_{d}$ & $C^{5_{2}}_{1}$ & $C^{5_{2}}_{1}$ & 
    $C^{5_{2}}_{2}$ & $C^{5_{2}}_{2}$ \\ \hline
     \vrule width 0pt height 13pt
   $F_{3} \sim Q_{3},L_{3}$ & $C^{5_{2}}_{2}$ & $C^{5_{2}}_{3}$ &
    $C^{5_{2}}_{3}$ & $C^{5_{2}}_{1}$ \\ \hline 
     \vrule width 0pt height 13pt
   $\bar{F}_{3} \sim U^{C}_{3} \, , \, D^{C}_{3} \, , \, E^{C}_{3} 
    \, , \, N^{C}_{3} $ & $C^{5_{2}}_{3}$ & $C^{5_{2}}_{2}$ & 
    $C^{5_{2}}_{1}$ & $C^{5_{2}}_{3}$ \\ \hline
  \end{tabular} 
   \caption {{\small Allocation of $5_{2}$ states that lead to third 
   family-only Yukawa couplings at lowest order.  We use the lower index 
   to distinguish between doublets, singlets and Higgs fields.}} 
  \label{tab:states1}
 \end{center}
\end{table}  

Notice that there are no free indices on the intersection states
$Q_{i} , L_{i} , U_{i}^{C} , D_{i}^{C} , E_{i}^{C} , N_{i}^{C} \equiv
C^{5_{1} 5_{2}}$ (i=1,2), which means that we cannot distinguish
between the first two families.

In our Type I string-inspired model, we shall assign the gauge groups 
and matter fields as in Table \ref{tab:states2}.  We ignore
the custodial $SU(4)_{5_{1}} \otimes U(1)^{6}$ symmetry.  The states
$\phi, \phi', H^{\alpha b}, \bar{H}_{\alpha b}$ are used to break the
gauge group down to the Standard Model as discussed in Appendix 
\ref{app:specgaugegen}.
\begin{table}[h]
 \begin{center}
  \begin{tabular}{|c|c||c|c|c|c|c|} \hline
   States & Sector & $SU(4)_{5_{2}}$ & $SU(2)_{5_{2R}}$ & $SU(2)_{5_{2L}}$ &
    $SU(2)_{5_{1R}}$ & $SU(2)_{5_{1L}}$ \\ \hline\hline 
     \vrule width 0pt height 13pt
    $F_{i} \sim Q_{i},L_{i}$ & $5_{1} 5_{2}$ & 4 & 1 & 1 & 1 & 2 \\ \hline
    \vrule width 0pt height 13pt
   $\bar{F}_{i} \sim U_{i}^{C},D_{i}^{C},E_{i}^{C},N_{i}^{C},$ &
    $5_{1} 5_{2}$ & $\bar{4}$ & 1 & 1 & 2 & 1 \\ \hline
     \vrule width 0pt height 13pt
    $F_{3} \sim Q_{3},L_{3}$ & $5_{2}$ & 4 & 1 & 2 & 1 & 1 \\ \hline
     \vrule width 0pt height 13pt
   $\bar{F}_{3} \sim U_{3}^{C},D_{3}^{C},E_{3}^{C},N_{3}^{C},$ &
    $5_{2}$ & $\bar{4}$ & 2 & 1 & 1 & 1 \\ \hline
     \vrule width 0pt height 13pt
   $\phi$ & $5_{1} 5_{2}$ & 1 & 1 & 2 & 1 & 2 \\ \hline
     \vrule width 0pt height 13pt
   $\phi'$ & $5_{1} 5_{2}$ & 1 & 2  & 1 & 2 & 1 \\ \hline 
     \vrule width 0pt height 13pt
   $H^{\alpha b}$ & $5_{2}$ & 4 & 2 & 1 & 1 & 1 \\ \hline
     \vrule width 0pt height 13pt
   $\bar{H}_{\alpha b}$ & $5_{2}$ & $\bar{4}$ & 2 & 1 & 1 & 1 \\ \hline
     \vrule width 0pt height 13pt
   $h \sim h_{u},h_{d}$ & $5_{2}$ & 1 & 2 & 2 & 1 & 1 \\ \hline 
 \end{tabular}
  \caption {{\small $SU(4)_{5_{2}} \otimes
  SU(2)_{5_{2R}} \otimes SU(2)_{5_{2L}} \otimes SU(2)_{5_{1R}} \otimes
  SU(2)_{5_{1L}}$ quantum numbers for left and right-handed chiral
  fermion states and symmetry breaking Higgs fields.}}
  \label{tab:states2} 
 \end{center}
\end{table}  

Gauge invariance with respect to the initial gauge group
$SU(4)_{5_{2}} \otimes SU(2)_{5_{2R}} \otimes SU(2)_{5_{2L}}
\otimes SU(2)_{5_{1R}} \otimes SU(2)_{5_{1L}}$ provides the mechanism
to forbid both first and second family Yukawa couplings 
$\left( \bar{F}_{i} F_{j} h \right)$ and R-parity violating operators
without any other assumptions\footnote{Note that the third family 
right-handed neutrinos and sneutrinos receive large Majorana masses from
the operators $\bar{F}_{3} \bar{F}_{3} H H$ resulting in a see-saw mechanism.
This is discussed in Ref. \cite{king}, along with a discussion of 
higher-dimensional operators suitable for first and second family fermion
masses.}.

Note that the $\mu$-term is forbidden by string selection rules which also
forbid a superpotential term involving a matter brane singlet\footnote{A
non-renormalisable higher-dimensional 4-point superpotential term may
be generated by two additional gauge singlet fields, eg. $W \sim N_{1}
N_{2} h_{u} h_{d}$.  This can become the 3-point term when one of the
singlet fields acquire a vev.} where ${\mbox W \sim \lambda N h_{u}
h_{d}}$.  The Giudice-Masiero mechanism offers the best opportunity
of producing a $\mu$-term from the soft potential as discussed later.

\section{Brane Mediated Supersymmetry Breaking}  \label{sec:bmsb}

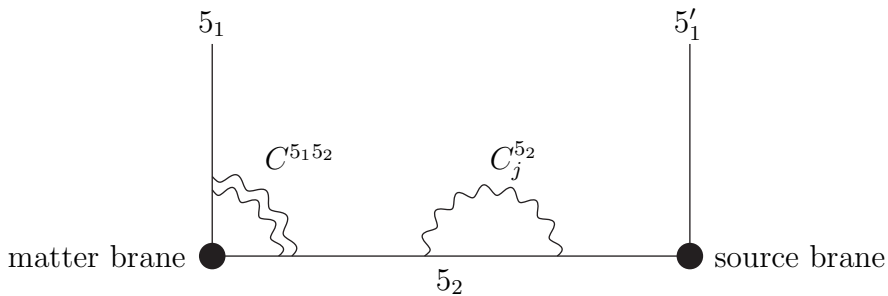
\begin{figure}[h]
  \begin{center} 
   \begin{picture}(190,100)(5,10)
    \Line(5,10)(5,90) 
    \Line(5,10)(185,10) 
    \Line(185,10)(185,90)
    \Vertex(5,10){5}    \Text(-5,10)[r]{matter brane}
    \Vertex(185,10){5}  \Text(195,10)[l]{source brane}
    \Text(5,95)[b]{$5_{1}$}
    \Text(185,95)[b]{$5_{1}'$}
    \Text(95,5)[t]{$5_{2}$}
    \PhotonArc(5,10)(25,0,90){2}{4} 
    \PhotonArc(5,10)(30,0,90){2}{4}
    \Text(110,43)[lb]{$C^{5_{2}}_{j} $} 
    \Text(25,43)[lb]{$C^{5_{1} 5_{2}} $} 
    \PhotonArc(110,10)(25,0,180){2}{8}
   \end{picture}
  \end{center} 
 \caption{{\small A brane-construction using overlapping D5-branes, 
  with effective D3-branes at the intersection points spatially 
  separated along the $D5_{2}$-brane.  The first two chiral families
  $(C^{5_{1} 5_{2}})$ live on the first intersection region.  The 
  third family and Higgs doublets $(C_{j}^{5_{2}})$ live on the
  $D5_{2}$-brane in the ``bulk'' between the source and matter branes.
  The gauge-singlet source field in principle can either live on the 
  $D5'_{1}$-brane or be localised on the $5'_{1} 5_{2}$ intersection,
  but for definiteness we assume the latter possibility.}}
  \label{fig:construct}
 \end{figure}

We now augment the model in section \ref{sec:typei},
including the states in Table \ref{tab:states2}, by including an
additional $D5_{1}'$-brane located at an orbifold fixed point away from 
the origin as shown in Fig. \ref{fig:construct}.
The idea of including the extra $5_{1}'$-brane is that SUSY is broken
on this brane and communicated by the MSSM states that live on the 
$5_{2}$-brane which intersects it.  Thus, the gauge fields on the 
$5_{2}$-brane play the role of the gauge fields in the bulk in Fig.
\ref{fig:loop}.  Note that there are many mass scales in this model 
as discussed in Appendix \ref{app:massscales}.

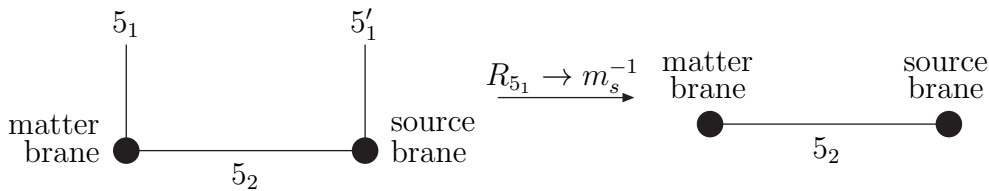
\begin{figure}[h] 
  \begin{center}
       \begin{picture}(320,60)(5,10) 
          \Line(5,10)(5,50)                 \Line(5,10)(95,10)
          \Line(95,10)(95,50)
          \Vertex(5,10){5}
          \Text(-5,10)[r]{brane}          \Text(-5,20)[r]{matter}
          \Vertex(95,10){5}                
          \Text(105,10)[l]{brane}         \Text(105,20)[l]{source}
          \Text(5,55)[b]{$5_{1}$}
          \Text(95,55)[b]{$5_{1}'$}
          \Text(50,5)[t]{$5_{2}$}
          \LongArrow(145,30)(195,30)
          \Text(170,35)[b]{$R_{5_{1}} \rightarrow m_{s}^{-1}$}
          \Line(225,20)(315,20)   \Text(270,15)[t]{$5_{2}$}
          \Vertex(225,20){5}      \Text(225,30)[b]{brane}
          \Text(225,40)[b]{matter}
          \Vertex(315,20){5}      \Text(315,30)[b]{brane} 
          \Text(315,40)[b]{source}
        \end{picture} 
  \end{center} 
 \caption{{\small The intersecting D5-brane construction in the limit 
 of small $D5_{1}$ compactification radius.  The $D5_{1}$-branes reduce 
 to effective D3-branes, separated in two orthogonal dimensions along 
 the $D5_{2}$-brane ({\it bulk}).  The allocation of Higgs and chiral 
 matter fields are the same as in Fig. \ref{fig:construct}
 and Table \ref{tab:states2}.}}
  \label{fig:asymm}
\end{figure}

We now consider a limiting case in which the model in Fig. 
\ref{fig:construct} reduces to the $\tilde{g}MSB$ model 
discussed in section \ref{sec:gauginomed}, namely that the D$5_{2}$
radius of compactification is very much larger than the $D5_{1}$ 
radius\footnote{Both inverse radii must be larger than the inverse 
string scale.}
\begin{equation}
  R_{5_{2}} \gg R_{5_{1}} \gg m_{s}^{-1}
\end{equation}
In this limit, the model reduces to that shown in Fig. \ref{fig:loop}, 
where the D3-branes correspond to the intersection regions of the D5-branes,
and the bulk corresponds to the mediating $5_2$ brane,
as shown in Fig. \ref{fig:asymm}. Note that the 
first two families are located on the matter brane, while
the third family and Higgs doublets live on the mediating $5_2$ brane. 

Since the gauge couplings on the branes are given by 
\begin{equation} 
  g_{5_{2}}^{-2} = \frac{ m^{2}_{s} v_{2} }{ (2\pi)^{3} \lambda },
  \hspace{8mm} g_{5_{1}}^{-2} = \frac{
  m^{2}_{s} v_{1} }{ (2\pi)^{3} \lambda } \label{eq:gaugecoupling}
\end{equation}
we know that the coupling-squared
is inversely proportional to compactification volume ($v_{i} \sim (2 \pi
R_{i})^{2}$), which implies that $g_{5_{1}} \gg g_{5_{2}}$.
This limiting case of the symmetry breaking is discussed in
Appendix \ref{app:specgaugelimit}, but the important results are that the
dominant components of the gauge fields live on the $D5_{2}$-brane 
which is consistent with $\tilde{g}MSB$ with two extra bulk
dimensions.  After the gauge symmetry is broken down to the Standard
Model, we recover the
relationship between gauge couplings:
\begin{equation}
  g'_{Y} \sim \sqrt{\frac{3}{5}} g_{5_{2}}
\end{equation}
(where $g_{3} \equiv g_{5_{2}}$).  This is consistent with gauge coupling
unification if $g_{5_{2}} \equiv g_{GUT}$ at the GUT scale.
 
It is also interesting to note that the restrictions we place on the
radii do not restrict the radius of the third complexified dimension 
too strongly.  This could allow a large extra dimension felt by gravity 
alone (with a size of the order of 1mm) as considered recently
\cite{largeexdim}, but we will not discuss that possibility here.

In this limiting case, we can use the results of Ref. \cite{chacko}
where we identify $L \equiv R_{5_{2}}$.  We can extend the analysis for 
the size of the extra dimensions and exponential suppression factors.
Ref. \cite{chacko} considers the maximum dimension size in the strong
coupling limit $\epsilon \sim 1$, but for a small number of extra 
dimensions, the theory does not need to be strongly coupled at the string 
scale, ie. $\epsilon \neq 1$.

Consider our symmetric toroidal compactification where the volume of the
compact dimensions is
\begin{equation}
   V_{5_{2}} \sim L^{2} \equiv R_{5_{2}}^{2}  \label{eq:volume}
\end{equation}
Using eqs. (\ref{eq:gscaling},\ref{eq:gdscaling}) with $D=6$, we can relate 
dimension size to $\epsilon$ for a 4-dimensional gauge coupling of order 1 
(as observed for SM couplings).
\begin{eqnarray}
  g_{5_{2}}^{2} \sim \frac{\epsilon l_{6}}{m_{s}^{2}} \sim L^{2} \nonumber \\
  L m_{s} \sim \left( \epsilon l_{6} \right)^{\frac{1}{2}}  \label{eq:g52}
\end{eqnarray}
Note that from eqs. (\ref{eq:volume}, \ref{eq:g52}), we have
\begin{equation}
  V_{5_{2}} m_{s}^{2} \sim \epsilon l_{6}  \label{eq:useful}
\end{equation}
\begin{table}[h]
 \begin{center}
  \begin{tabular}{|c||c|c|} \hline
   $\epsilon$ & $L m_{s}$ & $e^{-L m_{s}/2}$ \\ \hline\hline 
   $1$    & $63$ & $2 \times 10^{-14}$ \\ \hline 
   $0.8$  & $56$ & $6 \times 10^{-13}$ \\ \hline 
   $0.6$  & $49$ & $3 \times 10^{-11}$ \\ \hline 
   $0.4$  & $40$ & $2 \times 10^{-9}$  \\ \hline 
   $0.2$  & $28$ & $8 \times 10^{-7}$  \\ \hline 
   $0.1$  & $20$ & $5 \times 10^{-5}$  \\ \hline 
   $0.05$ & $14$ & $9 \times 10^{-4}$  \\ \hline
   $0.01$ & $6$  & $4 \times 10^{-2}$  \\ \hline
  \end{tabular}
   \caption {{\small Estimates for the toroidal compactification length L
    and exponential suppression factor for $D=6$, where $L \equiv R_{5_{2}}$.}}
     \label{tab:est2}
 \end{center}
\end{table}  

We have just seen how to recover the $\tilde{g}MSB$ model, but with two
extra dimensions and the third family in the bulk.  We can therefore use
the $\tilde{g}MSB$ results for the operators that lead to scalar and Higgs masses, A 
and $B\mu$-terms and even a $\mu$-term via the Giudice-Masiero
mechanism\footnote{Remember that a superpotential $\mu$-term is forbidden
by string selection rules for {\it our} choice of states.}.  However, in 
our model with $M \equiv m_{s}$ and 
${\mbox R_{5_{2}} \gg R_{5_{1}} > m_{s}^{-1} }$, there are only two extra
dimensions in the bulk between D3-branes\footnote{This allows us to use
Table \ref{tab:est2} to get restrictions on the size of $R_{5_{2}}$.}.

We use the eqs. (\ref{eq:contact},\ref{eq:branel1},\ref{eq:branel2})
with the following identifications:
\begin{eqnarray} 
 \phi_{M} &\equiv& \left( C^{5_{1} 5_{2}} \right) \hspace{5mm} Q_{i} , 
  L_{i} , U^{C}_{i} , D^{C}_{i} , E^{C}_{i} , N^{C}_{i} 
  \hspace{1cm} (i=1,2) \nonumber \\
 W^{\alpha} &\equiv& W_{SM} \nonumber \\ 
 h_{u} , h_{d} &\equiv& \left( C^{5_{2}}_{j} \right) \hspace{5mm} 
  h_{u} , h_{d} , Q_{3} , L_{3} , U^{C}_{3} , D^{C}_{3} , E^{C}_{3} , 
  N^{C}_{3} \\
 \hat{\phi}_{S} &\equiv& \left( C^{5_{1}' 5_{2}} \right) \hspace{5mm} S 
  \nonumber
\end{eqnarray}
to generate higher-dimensional operators, subject to the full 42222
gauge invariance.  We assume that the  F-component of the
gauge-singlet field S, which we assume to be an open string state
on the intersection between the souce brane and the mediating brane,
acquires a non-zero vev and breaks supersymmetry.  We now proceed to
discuss the different types of masses in the limiting case of the 
BMSB model.

\subsection{Gaugino masses}  \label{sec:gmass}

In the limit of $R_{5_{1}} < R_{5_{2}}$, the Standard Model gauge
fields are dominated by their components on the $D5_{2}$-brane (bulk).
In agreement with $\tilde{g}MSB$, we generate gaugino masses of the
same order of magnitude from eqs. (\ref{eq:branel1},\ref{eq:branel2})
\begin{equation}
  m_{\lambda} \sim \frac{ F_{S} }{m_{s}} \frac{ l_{6}/l_{4} }{ m_{s}^{2} 
   V_{ 5_{2} } } \sim \frac{1}{\epsilon l_{4}} \frac{ F_{S} }{m_{s}} 
    \label{eq:gmass2}
\end{equation}
(where $V_{5_{2}}$ is the volume of the compact dimensions inside the
$D5_{2}$-brane world-volume.)

\subsection{First and second family scalar masses}

This is the generic 4-point contact term between fields on opposite
branes that leads to exponentially suppressed first and second
family squark and slepton masses\footnote{This operator also leads to
first and second family mixing and off-diagonal mass matrix elements.
There may be another operator leading to first and third family
mixing, eg.  $\triangle {\mathcal L} \sim \int d^{4} \theta C^{\dagger
5_{1} 5_{2}} C^{5_{2}}_{j} S^{\dagger} S $.}, using eq.
(\ref{eq:contact}).

\begin{eqnarray} 
 \triangle {\mathcal L}_{soft} &\sim& 
  \frac{ e^{- m_{s} R_{5_{2}} } }{ m_{s}^{2} } \int d^{4} \theta
  C^{\dagger 5_{1} 5_{2}} C^{5_{1} 5_{2}} S^{\dagger} S \\
  V_{soft} &\sim& \frac{ e^{- m_{s} R_{5_{2}} } 
  F_{S}^{2} }{ m_{s}^{2} } \left( \tilde{Q}^{*}_{i} \tilde{Q}_{j} +
  \tilde{U}^{C *}_{i} \tilde{U}^{C}_{j} + \tilde{D}^{C *}_{i} 
  \tilde{D}^{C}_{j} + \tilde{L}^{*}_{i} \tilde{L}_{j} + \tilde{E}^{C *}_{i}
  \tilde{E}^{C}_{j} + \tilde{N}^{C *}_{i} \tilde{N}^{C}_{j} \right)  
      \label{eq:vsoft12}
\end{eqnarray}
Table (\ref{tab:est2}) shows that the exponential suppression factor
is strong for two extra dimensions.  Therefore, contact term contributions to
the first and second family scalar masses are negligible at high energies. 
Instead, they are generated by Renormalization Group Equation (RGE) effects.

Loop contributions to first and second family scalar masses (Fig.
\ref{fig:loop}) are much larger than contact terms and
anomaly mediated contributions.  So, although the first
and second family squark/slepton masses are not zero at high-energies,
they are suppressed by a loop factor relative to third family scalar
masses.

\subsection{Higgs mass terms and third family scalar masses}

Extending eq. (\ref{eq:branel1}) to include third family scalars, we have the
following higher-dimensional operators:

\begin{eqnarray} 
 \triangle {\mathcal L}_{soft} \sim \frac{ l_{6} }{ l_{4} } \int d^{4} 
  \theta \left\{ \frac{1}{ m_{s}^{3} } \left( S^{\dagger} h_{u} h_{d} +
  h.c. \right) + \frac{1}{ m_{s}^{4} } S^{\dagger} S 
  \left[ h_{u}^{\dagger} h_{u} + h_{d}^{\dagger} h_{d} \right. \right. 
  \nonumber \\ 
 \left. \left. + \left( h_{u} h_{d} + h.c. \right) + Q^{\dagger}_{3} Q_{3}
  + U^{C \dagger}_{3} U^{C}_{3} + D^{C \dagger}_{3} D^{C}_{3} + 
  L^{\dagger}_{3} L_{3} + E^{C \dagger}_{3} E^{C}_{3} + N^{C \dagger}_{3}
  N^{C}_{3} \right] \right\}  \label{eq:lsoft2}
\end{eqnarray}
From eqs. (\ref{eq:branel2}, \ref{eq:lsoft2}), we obtain the $\mu$-term,
\begin{equation}
    \mu \sim \frac{ F_{S} }{m_{s}} \frac{ l_{6}/l_{4} }{ m_{s}^{2} 
     V_{ 5_{2} } } \sim \frac{1}{\epsilon l_{4}} \frac{ F_{S} }{m_{s}} 
\end{equation}
Higgs and third family scalar masses.
\begin{equation}
   B\mu , m_{h_{u}}^{2} , m_{h_{d}}^{2} , m_{\tilde{F}_{3}}^{2} \sim 
    \frac{ F_{S}^{2} }{ m_{s}^{2} } \frac{ l_{6}/l_{4} }{ m_{s}^{2} 
     V_{5_{2}} } \sim \frac{1}{\epsilon l_{4}} \frac{ F_{S}^{2} }{m_{s}^{2}}
\end{equation}
\begin{center}
 (where $\tilde{F}_{3} \equiv \tilde{Q}_{3} , \tilde{U}^{C}_{3} ,
  \tilde{D}^{C}_{3} , \tilde{L}_{3} , \tilde{E}^{C}_{3} ,
  \tilde{N}^{C}_{3}$)
\end{center}

\subsection{Scalar mass matrix}

We have generated a scalar mass matrix
with an explicit third family mass hierarchy at lowest order:
\begin{equation}
  m_{scalar}^{2} \sim \frac{1}{\epsilon l_{4}} \frac{ F_{S}^{2} }{m_{s}^{2}}
   \left( 
 \begin{array}{ccc} 0 & 0 & 0 \\ 
                    0 & 0 & 0 \\
                    0 & 0 & 1 
 \end{array} \right)
\end{equation}
The first and second family mass matrix elements are dominated by loop
corrections since the contact term contributions are 
exponentially suppressed.  However these contributions are still
smaller than the third family masses due to the location of the third
family in the bulk and its direct coupling to the SUSY breaking
hidden sector.
 
\subsection{Trilinear A-terms}

Gauge invariant operators can be constructed for third family A-terms as 
follows:
\begin{equation}
  \triangle {\mathcal L}_{soft} \sim \frac{l_6}{l_4}
  \int d^{2} \theta \frac{1}{m_{s}^{4}} S
   \left( h_{d} D^{C}_{3} Q_{3} + h_{u} U^{C}_{3} Q_{3} + h_{d}
   E^{C}_{3} L_{3} + h_{u} N^{C}_{3} L_{3} \right) + h.c.
\end{equation}
These operators lead to trilinear A-terms:
\begin{equation}
A_{ij} \sim \frac{ F_{S} }{m_{s}} \frac{l_6/l_4}{m_{s}^{3} 
       V_{5_{2}}^{3/2}} \left(
 \begin{array}{ccc}
   0 & 0 & 0 \\
   0 & 0 & 0 \\
   0 & 0 & 1
 \end{array}
\right)
\sim \frac{ F_{S} }{m_{s}} 
 \frac{1}{\epsilon l_{4} \left(\epsilon l_{6} \right)^{1/2}} \left(
 \begin{array}{ccc}
   0 & 0 & 0 \\
   0 & 0 & 0 \\
   0 & 0 & 1
 \end{array}
\right)  \label{eq:aterm}
\end{equation}
using eq. (\ref{eq:useful}).

The first and second family A-terms are negligible in comparison to
the third family term.  Instead, they will receive loop-suppressed
contributions.

\subsection{Yukawa textures}

Using our choice of states and eq.
(\ref{eq:superp2}), we obtain a third family
hierarchical Yukawa texture for the quark and lepton sectors at
lowest-order.  This texture reflects the observation that 
${\mbox m_{t} \gg m_{c} , m_{u}}$ ; ${\mbox m_{b} \gg m_{s} , m_{d}}$ and 
${\mbox m_{\tau} > m_{\mu} , m_{e}}$.
\begin{equation}
Y^{a}_{ij} \sim
\left(
 \begin{array}{ccc}
   0 & 0 & 0 \\
   0 & 0 & 0 \\
   0 & 0 & 1
 \end{array}
\right) \,\, {\mathrm where } \,\, a \equiv u,d,e,n 
\end{equation}

Smaller NLO Yukawa couplings (and associated trilinear A-terms) are
generated by higher-dimensional operators.  Notice that an interesting 
operator is allowed by 42222 gauge invariance, and appears to be such a 
small Yukawa term:
\begin{equation}
   \triangle {\mathcal L} \sim \bar{F}_{i} F_{i} h \phi \phi'
\end{equation}

The fields $h$, $\phi$ and $\phi'$ (Higgs) acquire vevs and
spontaneously break the gauge symmetry.  When each field is replaced
by its vev, we can generate a first and second family mass term.  This
operator will be suppressed by powers of the string scale such that the 
first and second family have much smaller masses relative to 
the third family in the bulk.

\subsection{Mass ratios and FCNC constraints}

Consider the ratio of Higgs and third family scalar masses 
$B\mu , m_{h_{u}}^{2} , m_{h_{d}}^{2} ,
m_{\tilde{F}_{3}}^{2}$ to gaugino masses $m_{\lambda}^{2}$:
\begin{equation}
  \frac{m_{\phi}^{2}}{m_{\lambda}^{2}} \sim \frac{l_{4}}{l_{6}}
  m_{s}^{2} V_{5_{2}} \sim \epsilon l_{4}
\end{equation}
(using eqs. (\ref{eq:gscaling},\ref{eq:gdscaling}) and $g_{4}
\sim 1$, where ${\mbox m_{\phi}^{2} \equiv B\mu , m_{h_{u}}^{2} ,
m_{h_{d}}^{2} , m_{\tilde{F}_{3}}^{2}}$)

Also consider the ratio of trilinear soft masses $A_{33}$
to gaugino masses $m_{\lambda}$ using eqs. (\ref{eq:gmass2},\ref{eq:aterm}):
\begin{equation}
  \frac{A_{33}}{m_{\lambda}} \sim \frac{1}{\left( \epsilon {l_{6}}
     \right)^{1/2}}
\end{equation}
\begin{table}[h]
 \begin{center} 
  \begin{tabular}{|c||c|c|c|c|} \hline
   $\epsilon$ & $ e^{- L m_{s} / 2}$ & $m_{\phi}^{2}/m_{\lambda}^{2}$ 
    & $m_{\phi}/m_{\lambda}$ & $A_{33}/m_{\lambda}$ \\ \hline\hline 
   1.0  & $2\times10^{-14}$ & 158 & 12.6 & 0.016 \\ \hline 
   0.8  & $6\times10^{-13}$ & 126 & 11.2 & 0.018 \\ \hline
   0.6  & $3\times10^{-11}$ &  95 & 9.7  & 0.020 \\ \hline 
   0.4  & $2\times10^{-9}$  &  63 & 7.9  & 0.025 \\ \hline
   0.2  & $8\times10^{-7}$  &  32 & 5.6  & 0.035 \\ \hline
   0.1  & $5\times10^{-5}$  &  16 & 4.0  & 0.050 \\ \hline 
   0.05 & $9\times10^{-4}$  &   8 & 2.8  & 0.071 \\ \hline 
   0.01 & $4\times10^{-2}$  & 1.6 & 1.3  & 0.159 \\ \hline 
  \end{tabular}
   \caption {{\small Estimates for the ratio of scalar masses and third 
             family A-terms to gaugino masses for different $\epsilon$ and 
             the exponential suppression factor (for masses-squared) arising 
             from toroidal compactification.}} 
  \label{tab:est3}
 \end{center}
\end{table}  

Experimental constraints on FCNC\footnote{See \cite{chacko}
and references therein.} from mass-squared matrix elements require an
exponential suppression of $\sim 10^{-3} - 10^{-4}$ for first and second
family scalar masses in eq. (\ref{eq:vsoft12}).  Using Table
\ref{tab:est3}, we get a {\it lower} limit of say $\epsilon \sim 0.01$.
However, phenomenological considerations restrict the ratio of
$m_{\phi}$ and $m_{\lambda}$, and places an {\it upper} limit of say
$\epsilon \sim 0.1$.  This amount of suppression requires that the effective
D3-branes are separated by a distance of order $\sim 10/m_{s}$.

\subsection{Phenomenology}

As in \cite{chacko} we shall consider the phenomenology 
based on an inverse compactification scale ($R_{5_2}^{-1}$ in our case)
close to the unification
scale $M_{GUT} \sim 2 \times 10^{16}$ GeV.
It is natural to assume a high energy unification scale 
in the limiting case  
$g_{5_{1}} \gg g_{5_{2}}$ since in this limit
the light physical gauge fields
all arise from the mediating $5_2$ brane, and so are
all subject to a single gauge coupling constant,
$g_{5_{2}}\equiv g_{GUT}$.

We have seen that in the BMSB model (at $M_{GUT}$)
the trilinear and first and second family soft
masses are negligible, while the third family 
soft masses, and the Higgs mass parameters are larger than the
gaugino masses.
In Table 5 we compare a sample spectrum in the
BMSB model to that in both the $\tilde{g}MSB$ model and the
no-scale supergravity model, where the ratio of Higgs
vevs $\tan \beta =20$ and a universal gaugino mass
of $M_{1/2}=300$ GeV are chosen to give a lightest Higgs boson mass
of about 115 GeV, consistent with the recent LEP signal \cite{LEP,Higgs}.
\footnote{Note that $\tan \beta =20$
is sufficiently small that we
may neglect all Yukawa couplings except the top Yukawa coupling
in the RGEs.}

\begin{table}[tbp]
\hfil
\begin{tabular}{cccc}
\hline
                         & BMSB & $\tilde{g}MSB$ & no-scale  
\\ \hline
$M_{1/2}$                & 300  & 300            & 300  
\\ \hline
$A_0$                    & 0    & 0              & 0
\\ \hline
$m_{\tilde{F}_{1,2}}$    & 0    & 0              & 0
\\ \hline
$m_{\tilde{F}_{3}}$      & 500  & 0              & 0
\\ \hline
$m_{h_{u}}$              & 500  & 500            & 0  
\\ \hline
$m_{h_{d}}$              & 500  & 500            & 0  
\\ \hline
$\tilde{g}$              & 830  & 830            & 830  
\\ \hline
$\tilde{\chi}_1^0$       & 124  & 119            & 124   
\\ \hline
$\tilde{\chi}_2^0$       & 239  & 200            &  237  
\\ \hline
$\tilde{\chi}_3^0$       & 506  & 258            &  472 
\\ \hline
$\tilde{\chi}_4^0$       & 517  & 314            &  485 
\\ \hline
$\tilde{\chi}_1^\pm$     & 238  & 195            &  237  
\\ \hline
$\tilde{\chi}_2^\pm$     & 518  & 314            &  486  
\\ \hline
$\tilde{E}_{L_{1,2}}$    & 220  & 220            &  220  
\\ \hline
$\tilde{E}_{L_3}$        & 546  & 220            &  220  
\\ \hline
$\tilde{E}_{R_{1,2}}$    & 124  & 124            &  124    
\\ \hline
$\tilde{E}_{R_3}$        & 515  & 124            &  124    
\\ \hline
$\tilde{N}_{L_{1,2}}$    & 205  & 205            &  205   
\\ \hline
$\tilde{N}_{L_3}$        & 540  & 205            &  205   
\\ \hline
$\tilde{U}_{L_{1,2}}$    & 740  & 740            &  740   
\\ \hline
$\tilde{U}_{L_3}$        & 783  & 653            &  676   
\\ \hline
$\tilde{U}_{R_{1,2}}$    & 715  & 715            &  715   
\\ \hline
$\tilde{U}_{R_3}$        & 628  & 520            &  577   
\\ \hline
$\tilde{D}_{L_{1,2}}$    & 744  & 744            &  744   
\\ \hline
$\tilde{D}_{L_3}$        & 787  & 658            &  681     
\\ \hline
$\tilde{D}_{R_{1,2}}$    & 713  & 713            &  713 
\\ \hline
$\tilde{D}_{R_3}$        & 871  & 713            &  713 
\\ \hline
$\tilde{t}_1$            & 613  & 492            &  544   
\\ \hline
$\tilde{t}_2$            & 832  & 718            &  745  
\\ \hline
$\tan \beta$             & 20   & 20             &  20  
\\ \hline
$m_{h^0}$                & 115  & 114            &  115  
\\ \hline
$m_{H^0}$                & 738  & 596            &  511  
\\ \hline
$m_A$                    & 738  & 596            &  511 
\\ \hline
$m_{H^{\pm}}$            & 742  & 602            &  517  
\\ \hline
$\mu(M_Z)$             & 500    & 250            &  467 
\\ \hline
\end{tabular}
\hfil
\label{spectrum}
\caption{\small Comparison of spectra (in GeV) for the three models
BMSB, $\tilde{g}MSB$ and no-scale supergravity.
The common parameters are $\tan \beta =20$, universal gaugino mass
$M_{1/2}=300$ GeV, trilinear soft mass $A_0=0$, first and second
family squark and slepton masses $m_{\tilde{F}_{1,2}}^{2}=0$.
The parameters are chosen to give a lightest Higgs boson mass consistent
with the LEP signal \cite{LEP,Higgs}.
The $\mu$ parameter (assumed positive) and B are determined
from the low energy electroweak symmetry breaking conditions.}
\end{table}

In the no-scale model the only non-zero soft mass is $M_{1/2}$,
which results in a very characteristic spectrum where the
right-handed slepton is very light and is in danger of becoming
lighter than the lightest neutralino. The $\tilde{g}MSB$ model
differs from the no-scale model only by the inclusion of Higgs
soft masses which we have taken to be degenerate and somewhat
higher than the gaugino masses. The main effect is to reduce the
$\mu$ parameter, which is determined here from the electroweak symmetry
breaking condition, and taken to be positive, which results in
lighter charginos and neutralinos. Also in the $\tilde{g}MSB$ model
the heavy Higgs and third family squark spectrum is also
noticeably different from the no-scale model. 
\footnote{As noted in \cite{chacko}, if we had taken non-degenerate Higgs
soft masses then the lightest right-handed slepton mass could have
been significantly increased relative to the no-scale model
due to the hypercharge Fayet-Illiopoulos
term.} Turning to the BMSB model, we see that the effect
of having both the Higgs and third family soft masses is to raise the 
$\mu$ parameter, and of course to significantly increase the
third family squark and slepton masses, providing an unmistakable
spectrum and a characteristic smoking gun signature of the model.

\section{Conclusions}  \label{sec:conc}

We have proposed a mechanism for mediating SUSY breaking
in Type I string theories - BMSB. Rather similar to the
$\tilde{g}MSB$ set-up in Fig.\ref{fig:loop}
we have proposed a Type I string-inspired set-up consisting
of three intersecting D5-branes as shown in Fig.\ref{fig:construct}
in which the gauge fields, Higgs doublets and third family matter
fields all live on the third mediating 
$5_2$-brane which plays the role of the bulk in the $\tilde{g}MSB$ scenario.
The presence of the third matter family on the mediating D-brane
is characteristic of Type I string constructions and
provides the main experimentally testable difference between 
the BMSB and $\tilde{g}MSB$ models.

We have considered a limiting case
in which $R_{5_2}\gg R_{5_1}$, and shown that in this case
the model reduces to the original $\tilde{g}MSB$ model
with the role of the bulk being played by the mediating $5_2$-brane.
In this limiting case, 
the model naturally leads to approximately universal gaugino masses
and a single unified gauge coupling constant, which 
motivates the identification of the string scale with
the usual GUT scale. In this case
the phenomenology of the BMSB model is rather interesting,
and it may be compared to the predictions of the
no-scale supergravity and the $\tilde{g}MSB$ model.
As in the $\tilde{g}MSB$ model,
the first two families naturally receive very small masses at the
high energy scale leading to 
flavour-changing neutral currents being naturally suppressed. 
The presence of third family soft
masses will not alter this conclusion very much since
FCNC limits involving the third family are much weaker. 
However the third family soft masses will lead to 
a characteristic squark and slepton mass spectrum 
which may be easily distinguished from that of both
no-scale supergravity and the $\tilde{g}MSB$ model as
shown in Table 5.
The $\mu$-problem is solved by the Giudice-Masiero mechanism as in
the original $\tilde{g}MSB$ model.

In this limiting case the BMSB model bears a close resemblance to
both the no-scale supergravity and the $\tilde{g}MSB$ models.
The fact that the third family receives a non-zero soft 
SUSY breaking mass is strictly not an unambiguous signal
of the underlying Type I string model, since it is possible 
for this to happen in both the other cases also.
For example in the old heterotic models based on orbifolds,
matter may be localised in the fixed points of the orbifold
(the twisted sector) or not (the untwisted sector) so it is
possible to have the third family playing a different
role from that of the first two families. What is different
in the model presented here is that the gauge group
is localised on two different branes, but in the limiting case
(above) the physical gauge group arises essentially from one brane,
and in this limit we return to a situation similar to that 
of the old heterotic string theories. There are however three points
worth noting here. Firstly, the presence of two families
at the intersection points of two branes, and one family 
on a single brane, seems to be typical of Type I string 
constructions \cite{shiutye}. Secondly in Type I string constructions
we have the possiblity of full unification of both gravity and
gauge forces, precisely because gravity exists in 10 dimensions
whereas the gauge groups live in a 6 dimensional sub-manifold,
which is not possible in old heterotic string theories.
Thirdly, the limiting case of $R_{5_2}\gg R_{5_1}$ would be
expected to apply only approximately, and therefore in practice there will
be corrections, for example, to gauge
coupling unification which may be observable. 
Further comments concerning the non-limiting case are briefly
discussed below.

In the more general non-limiting case, the model will
have an even richer structure.
In this non-extremal radii limit
(ie. ${\mbox R_{5_{2}} > R_{5_{1}} \neq m_{s}^{-1}}$), we must use the full
gauge state expressions listed in Appendix \ref{app:specgaugegen}.  The
light gauge
states are no longer dominated by their $D5_{2}$-brane components, but
are instead mixtures of fields from either brane, with the exception of the
gluon/gluino states that {\it only} arise from the $D5_{2}$-brane.  
The result is 
that the high energy gluino mass will be larger than the 
high energy wino and bino masses.
In this more general case the gauge couplings are no longer
equal, so there is less motivation to identify the 
string scale with the GUT scale. Generally the string scale
can take any value from a few TeV to $10^{16}$ GeV, and we have the
possibility of a mm scale large extra dimension.  

The toy model has other interesting features such as the fact that
the gauge symmetry forbids first and second family Yukawa couplings
at lowest order, and naturally forbids R-parity violating operators that
cannot be forbidden by string selection rules alone, while
allowing the third family Yukawa coupling. Most importantly, however, the
toy model demonstrates the BMSB mechanism, which is based on having
at least three branes with two different intersection points.
This minimum requirement
implies that constructions with all the branes at the origin fixed point
are inadequate for our purpose. Although
there are examples in the literature of intersecting branes at
different fixed points \cite{Wang}, such models
are generally more complicated than the simple set-up
considered here. Nevertheless our BMSB mechanism could provide
a useful alternative starting point from which to address the
problem of SUSY breaking in more general Type I string theories.
   
Finally note that it has been been suggested,
in the context of Type I theories, that singlet twisted moduli,
which appear in the tree-level gauge kinetic function, might be
responsible for generating gaugino masses if they acquire
non-vanishing F-terms, and that this might provide a brane
realisation of $\tilde{g}MSB$ if the standard model gauge symmetry
originates from 9-branes providing that there are in addition
two sets of D5-branes located at two different fixed points \cite{Benakli}.
This suggestion shares some of the features with the present paper,
although model building issues were not discussed, and the 
characteristic possibility
of the third family on the mediating brane was not considered.
Also additional contributions from the F-terms of dilaton $S$ and 
moduli fields $T_i$ were also generically allowed, whereas here
we have implicitly assumed them to be absent.

\newpage
\begin{center}
{\bf \large Acknowledgements}
\end{center}
S.K. and D.R. would like to thank PPARC for a Senior Fellowship
and a Studentship.
S.K. also acknowledges very illuminating
discussions with Z.Chacko and Jing Wang at
the Aspen Center for Physics and is particularly indebted to
Lisa Everett and Gordy Kane 
at the University of Michigan for detailed discussions concerning
D-brane models and to Lisa Everett especially for
sharing her expertise with me.
We are also grateful to K.Benakli for pointing out the
existence of \cite{Benakli} which we were ignorant
of until after our paper was submitted
to the archive.

\appendix

\section{Spectrum of Gauge bosons}  
\subsection{General case}  \label{app:specgaugegen}

In this appendix we consider the effect of symmetry breaking on
massless gauge field states and gauge couplings. 
We begin with the gauge group $SU(4)_{5_{2}} \otimes
 SU(2)_{5_{1L}} \otimes SU(2)_{5_{2L}} \otimes SU(2)_{5_{1R}} \otimes
 SU(2)_{5_{2R}}$.  The couplings run with energy scales subject to
 RGEs.  Conventionally, the
 symmetry breaking all occurs at high energies $(10^{15}-10^{16} GeV)$
 except for ${\mbox SU(2)_{L} \otimes U(1)_{Y} \longrightarrow
 U(1)_{EM} }$ which happens at the Electroweak scale.  In the tables
 that follow, gauge couplings are assumed to be at high energies
 unless otherwise stated.  Notice that i, a and m are adjoint indices
 for SU(2), SU(3) and SU(4) respectively.
\begin{table}[h]
 \begin{center} \begin{tabular}{|c||c|c|c|c|c|} \hline\hline
  {\it Gauge group} & $SU(4)_{5_{2}}$ & $SU(2)_{5_{1L}}$ & $SU(2)_{5_{2L}}$
  & $SU(2)_{5_{1R}}$ & $SU(2)_{5_{2R}}$ \\ \hline
   {\it Coupling} & $g_{5_{2}}$ & $g_{5_{1}}$ & $g_{5_{2}}$ & $g_{5_{1}}$ & 
    $g_{5_{2}}$ \\
    & & & & & \\ 
   {\it States} & $G^{m}_{5_{2}}$ & $W^{i}_{5_{1}L}$ &
  $W^{i}_{5_{2}L}$ & $W^{i}_{5_{1}R}$ & $W^{i}_{5_{2}R}$ \\ \hline\hline
 \end{tabular}
 \caption {{\small The initial gauge groups, gauge couplings and states
 in our model.}}
   \end{center}
\end{table}

(a) First combine the chiral SU(2) groups from either brane via diagonal 
symmetry breaking to recover the Pati-Salam gauge group.
\begin{eqnarray}
  SU(2)_{5_{1L/R}} \otimes SU(2)_{5_{2L/R}}
    \stackrel{diagonal}{\stackrel{v_{\phi},v_{\phi'}}{\longrightarrow}}
    SU(2)_{L/R} \nonumber \\
  \Rightarrow SU(4)_{5_{2}} \otimes SU(2)_{L} \otimes SU(2)_{R} \equiv G_{PS}
\end{eqnarray}
Spontaneous symmetry breaking (SSB) induces a change of basis,
parametrised by
\begin{equation}
  \cos \theta_{\phi} = \frac{ g_{5_{2}} }{ \sqrt{ g_{5_{1}}^{2} 
 + g_{5_{2}}^{2} }} 
\end{equation}
We can express the new massless states and gauge couplings in terms of the 
original parameters.  The Higgs mechanism generates massive gauge bosons with 
masses of the order of the symmetry breaking scale.
\begin{table}[h]
 \begin{center}
  \begin{tabular}{|c||c|c|} \hline\hline 
   {\it Gauge group} & $SU(4)_{5_{2}}$ & $SU(2)_{L}$ 
    \hspace{25mm} $SU(2)_{R}$ \\ \hline
   {\it Coupling} & $g_{5_{2}}$ & ${\displaystyle 
    g_{L}=\frac{g_{5_{1}} g_{5_{2}} }{ \sqrt{ g_{5_{1}}^{2} + g_{5_{2}}^{2} } }
    = g_{R} }$ \\
    & &  \\
   {\it States} & $G^{m}_{5_{2}}$ & ${\displaystyle W^{i}_{L/R} = 
    \frac{1}{ \sqrt{ g_{5_{1}}^{2} + g_{5_{2}}^{2} } } \left( g_{5_{1}}
    W^{i}_{5_{2}L/R} + g_{5_{2}} W^{i}_{5_{1}L/R} \right) }$ \\ \hline\hline
  \end{tabular}
 \caption {{\small The new massless states and couplings after the original
 gauge symmetry is broken down to the Pati-Salam gauge group.}}
\end{center}
\end{table}
\begin{center}
 plus 3 massive $SU(2)_{L}$ $\left( \bar{W}_{L} \right)$ and 3
 massive $SU(2)_{R}$ $\left( \bar{W}_{R} \right)$ bosons, of mass
 $M_{\bar{W}_{L/R}}^{2} = \frac{1}{2} v_{\phi}^{2} \left(
 g_{5_{1}}^{2} + g_{5_{2}}^{2} \right) $
\end{center}
 
(b) QCD $SU(3)_{C}$ is contained within $SU(4)_{5_{2}}$. The U(1)s 
combine to give the Hypercharge U(1) using the relationship $Y= (B-L) + 2I_{R}$.
\begin{eqnarray}
   SU(4)_{5_{2}} \supset SU(3)_{C} \otimes U(1)_{B-L} \nonumber \\ 
   SU(2)_{R} \supset U(1)_{I_{R}} 
\end{eqnarray}
The Pati-Salam gauge group is broken down to the Standard Model by giving a vev
to a Higgs field H.
\begin{eqnarray}
   U(1)_{B-L} \otimes U(1)_{I_{R}} \stackrel{v_{H}}{\longrightarrow}
    U(1)_{Y} \nonumber \\ 
   \Rightarrow SU(3)_{C} \otimes SU(2)_{L} \otimes U(1)_{Y}
\end{eqnarray}
The change of basis is parametrised by
\begin{equation}
 \cos \theta_{H} = \frac{ {\sqrt \frac{3}{2}} g_{5_{2}}
}{ \sqrt{ g_{R}^{2} + \frac{3}{2} g_{5_{2}}^{2} } } = \sqrt{ \frac{ 3
\left( g_{5_{1}}^{2} + g_{5_{2}}^{2} \right) }{ 5 g_{5_{1}}^{2} + 3
g_{5_{2}}^{2} }} 
\end{equation}
\begin{table}[h]
 \begin{center}
 \scalebox{0.9}{
  \begin{tabular}{|c||c|c|c|} \hline\hline 
   {\it Gauge group} & $SU(3)_{C}$ & $SU(2)_{L}$ & $U(1)_{Y}$ \\ \hline 
   {\it Coupling} & $g_{5_{2}}$ & ${\displaystyle g_{L}=\frac{ g_{5_{1}}
    g_{5_{2}} }{ \sqrt{ g_{5_{1}}^{2} + g_{5_{2}}^{2} } }}$ &
    ${\displaystyle g^{'}_{Y}=\frac{ g_{5_{1}} g_{5_{2}} \sqrt{3} }{
    \sqrt{ 5 g_{5_{1}}^{2} + 3 g_{5_{2}}^{2} } }}$ \\
   & & & \\
   {\it States} & $G^{a}_{5_{2}}$ & ${\displaystyle W^{i}_{L} = \frac{
    g_{5_{1}} W^{i}_{5_{2}L} + g_{5_{2}} W^{i}_{5_{1}L} }{ \sqrt{
    g_{5_{1}}^{2} + g_{5_{2}}^{2} } }}$ & ${\displaystyle B_{Y} = \frac{
    \sqrt{ 3 } \left( g_{5_{1}} W^{3}_{5_{2R}} + g_{5_{2}}
    W^{3}_{5_{1R}} \right) + \sqrt{2} g_{5_{1}} G^{15}_{5_{2}} }{ \sqrt{
    5 g_{5_{1}}^{2} + 3 g_{5_{2}}^{2} } } }$ \\ \hline\hline
  \end{tabular}
 } 
   \caption {{\small The Standard Model massless states and gauge couplings
             expressed in terms of the original parameters.}} 
 \end{center}
\end{table}
\begin{center}
 plus 6 massive $SU(4)_{5_{2}}$ bosons $\left( G^{9}_{5_{2}} -
   G^{14}_{5_{2}} \right)$, mass $M_{G}^{2} = \frac{1}{4} v_{H}^{2}
   g_{5_{2}}^{2}$, \\ 2 massive $SU(2)_{R}$ bosons $\left( W^{\pm}_{R}
   \right)$, mass $M_{W^{\pm}_{R}}^{2} = \frac{1}{4} v_{H}^{2}
   g_{R}^{2}$, \\ and 1 massive $SU(2)_{B-L}$ boson $\left( X_{B-L}
   \right)$, mass $M_{X_{B-L}}^{2} = \frac{1}{4} v_{H}^{2} \left(
   g_{R}^{2} + \frac{3}{2} g_{5_{2}}^{2} \right) $
\end{center}
(c) Finally, we can recover the QCD and EM Standard Model gauge group via
the familiar low-energy Higgs mechanism, parametrised by
\begin{center}
${\displaystyle \cos \theta_{W} = \frac{ g_{L}(v_{h}) }{ \sqrt{
g_{L}^{2}(v_{h}) + g^{'2}_{Y}(v_{h}) } } = \sqrt{ \frac{ 5
g_{5_{1}}^{2}(v_{h}) + 3 g_{5_{2}}^{2}(v_{h}) }{ 8
g_{5_{1}}^{2}(v_{h}) + 6 g_{5_{2}}^{2}(v_{h}) }} }$.
\end{center}
Electroweak symmetry breaking occurs when the Higgs field h acquires a
non-zero vev.
\begin{eqnarray}
   SU(2)_{L} \otimes U(1)_{Y} \stackrel{v_{h}}{\longrightarrow}
    U(1)_{EM} \nonumber \\
   \Rightarrow SU(3)_{C} \otimes U(1)_{EM}
\end{eqnarray}
\begin{table}[h]
 \begin{center}
 \scalebox{0.9}{
  \begin{tabular}{|c||c|c|} \hline\hline 
   {\it Gauge group} & $SU(3)_{C}$ & $U(1)_{EM}$ \\ \hline 
   {\it Coupling} & $g_{5_{2}}(v_{h})$ & ${\displaystyle e = 
    \frac{ g_{5_{1}}(v_{h}) g_{5_{2}}(v_{h}) \sqrt{3} }{ \sqrt{ 8 
    g_{5_{1}}^{2}(v_{h}) + 6 g_{5_{2}}^{2}(v_{h}) } }}$ \\
   & & \\
   {\it States} & $G^{a}_{5_{2}}$ & ${\displaystyle A = \frac{ \sqrt{3} 
    g_{5_{1}}(v_{h}) \left( W^{3}_{5_{2}L} + W^{3}_{5_{2}R} \right) + 
    \sqrt{3} g_{5_{2}}(v_{h}) \left( W^{3}_{5_{1}L} + W^{3}_{5_{1}R} \right)
    + \sqrt{2} g_{5_{1}}(v_{h}) G^{15}_{5_{2}} }{ \sqrt{ 8 g_{5_{1}}^{2}(v_{h})
    + 6 g_{5_{2}}^{2}(v_{h}) } }}$ \\ \hline\hline 
  \end{tabular}
 }
   \caption {{\small The massless gauge states and couplings after electroweak
               symmetry breaking.}}
 \end{center}
\end{table}
\begin{center}
 plus 3 massive $SU(2)_{L}$ bosons $(W^{\pm}_{L}, Z^{0}_{L})$ with masses: \\ 
 ${\displaystyle M_{W^{\pm}_{L}} = \frac{1}{2} v_{h} g_{L}(v_{h}) = 
  \frac{ g_{5_{1}}(v_{h}) g_{5_{2}}(v_{h}) v_{h} }{ 2 
  \sqrt{ g_{5_{1}}^{2}(v_{h}) +g_{5_{2}}^{2}(v_{h}) }} }$ and \\
\vspace*{0.5cm}    
  ${\displaystyle M_{Z^{0}_{L}} = g_{5_{1}}(v_{h}) g_{5_{2}}(v_{h})
  v_{h} \sqrt{ \frac{ 4 g_{5_{1}}^{2}(v_{h}) + 3 g_{5_{2}}^{2}(v_{h})
  }{ 2 \left( g_{5_{1}}^{2}(v_{h}) + g_{5_{2}}^{2}(v_{h}) \right)
  \left( 5 g_{5_{1}}^{2}(v_{h}) + 3 g_{5_{2}}^{2}(v_{h}) \right) }} }$
\end{center}

\subsection{Limiting case $R_{5_{2}} \gg R_{5_{1}}$}  \label{app:specgaugelimit}

In this appendix we repeat the symmetry breaking analysis for the limiting 
case
\begin{equation}
  R_{5_{2}} \gg R_{5_{1}} \Leftrightarrow g_{5_{2}} \ll
  g_{5_{1}}
\end{equation}
We find that the dominant components of the massless gauge fields
live on the $5_{2}$-brane (``bulk'') which is consistent with $\tilde{g}MSB$.
 
(a) After diagonal symmetry breaking we recover the Pati-Salam gauge group

${\mbox SU(4)_{5_{2}} \otimes SU(2)_{L} \otimes SU(2)_{R} }$
\begin{table}[h]
 \begin{center}
  \begin{tabular}{|c||c|c|} \hline\hline 
   {\it Gauge group} & $SU(4)_{5_{2}}$ & $SU(2)_{L}$ \hspace{8mm} $SU(2)_{R}$
    \\ \hline 
   {\it Coupling} & $g_{5_{2}}$ & $g_{L/R} \sim g_{5_{2}}$ \\
    & & \\
   {\it States} & $G^{m}_{5_{2}}$ & $W^{i}_{L/R} \sim W^{i}_{5_{2}L/R}$ 
    \\ \hline\hline
  \end{tabular} 
   \caption {{\small The dominant components of massless states and couplings
               after symmetry has been broken down to the Pati-Salam group.}}
 \end{center}
\end{table}
\begin{center}
    plus 3 massive $SU(2)_{L}$ and 3 massive $SU(2)_{R}$ bosons
    $\left( \bar{W}_{L/R} \sim W^{i}_{5_{1} L/R} \right)$ , \\
    $M_{\bar{W}_{L/R}}^{2} \approx \frac{1}{2} v_{\phi}^{2}
    g_{5_{1}}^{2}$
\end{center}
(b) We break the Pati-Salam group down to the Standard Model.  Notice the 
relationship between the Hypercharge gauge coupling and the other gauge 
couplings, which is consistent with gauge coupling unification. This will
happen if the $5_{2}$ gauge coupling equals $g_{GUT}$ at the GUT scale.
\begin{table}[h]
 \begin{center} 
  \begin{tabular}{|c||c|c|c|} \hline\hline
   {\it Gauge group} & $SU(3)_{C}$ & $SU(2)_{L}$ & $U(1)_{Y}$ \\ \hline
   {\it Coupling} & $g_{5_{2}}$ & $g_{L} \sim g_{5_{2}}$ & $g^{'}_{Y} \sim
    \sqrt{\frac{3}{5}} g_{5_{2}}$ \\ 
   & & & \\
   {\it States} & $G^{a}_{5_{2}}$ & $W^{i}_{L} \sim W^{i}_{5_{2}L}$ & 
    $B_{Y} \sim \sqrt{\frac{3}{5}} W^{3}_{5_{2R}} + \sqrt{\frac{2}{5}} 
    G^{15}_{5_{2}}$ \\ \hline\hline
  \end{tabular}
   \caption {{\small The dominant components of the massless states and 
               couplings after the Pati-Salam group is broken down to the 
               Standard Model.}}
 \end{center}
\end{table}
  
\begin{center}
 plus 6 massive $SU(4)_{5_{2}}$ bosons $\left( G^{9}_{5_{2}} -
   G^{14}_{5_{2}} \right)$, $M_{G}^{2} \approx \frac{1}{4} v_{H}^{2}
   g_{5_{2}}^{2} $, \\ 
\hspace*{0.5cm}   
   2 massive $SU(2)_{R}$ bosons $\left(
   W^{\pm}_{R} \sim \frac{1}{\sqrt{2}} \left( W^{1}_{5_{2}R} \mp
   iW^{2}_{5_{2}R} \right) \right)$, $M_{W^{\pm}_{R}}^{2} \approx
   \frac{1}{4} v_{H}^{2} g_{5_{2}}^{2} $ \\
\hspace*{0.5cm}    and 1 massive
   $SU(2)_{B-L}$ boson $\left( X_{B-L} \sim \sqrt{\frac{3}{5}}
   G^{15}_{5_{2}} - \sqrt{\frac{2}{5}} W^{3}_{5_{2R}} \right)$,
   $M_{X_{B-L}}^{2} \approx \frac{5}{8} g_{5_{2}}^{2} v_{H}^{2}$
\end{center}
(c) Finally the Higgs mechanism induces electroweak symmetry breaking, and 
generates the massive W and Z bosons.
\begin{table}[h]
 \begin{center} 
  \begin{tabular}{|c||c|c|} \hline\hline 
   {\it Gauge group} & $SU(3)_{C}$ & $U(1)_{EM}$ \\ \hline
   {\it Coupling} & $g_{5_{2}}(v_{h})$ & $e \sim \sqrt{\frac{3}{8}}
    g_{5_{2}}(v_{h})$ \\
   & & \\
   {\it States} & $G^{a}_{5_{2}}$ & $A \sim \sqrt{\frac{3}{8}} 
    \left( W^{3}_{5_{2}L} + W^{3}_{5_{2}R} \right) + \frac{1}{2} G^{15}_{5_{2}}$ 
    \\ \hline\hline 
  \end{tabular}
  \caption {{\small The dominant components of the familiar massless gauge 
              states after electroweak symmetry.}}
 \end{center}
\end{table}
\begin{center}
 plus 3 massive $SU(2)_{L}$ bosons: \\ 
 $\left( W^{\pm}_{L} \sim \frac{1}{\sqrt{2}} \left( W^{1}_{5_{2}L} \mp 
  iW^{2}_{5_{2}L} \right) \right)$, $M_{W^{\pm}_{L}} \approx \frac{1}{2}
  v_{h} g_{5_{2}}(v_{h})$ \\ 
\hspace*{0.5cm}
  and $\left( Z^{0}_{L} \sim \sqrt{\frac{5}{8}} W^{3}_{5_{2}L} - 
   \frac{3}{2\sqrt{10}} W^{3}_{5_{2}R} - \frac{1}{2} \sqrt{\frac{3}{5}} 
   G^{15}_{5_{2}} \right)$, $M_{Z^{0}_{L}} \approx \sqrt{\frac{2}{5}} v_{h} 
   g_{5_{2}}(v_{h})$
\end{center}

\section{Mass scales}  \label{app:massscales}

In this appendix we consider the different mass scales present in the model.
Each time the gauge symmetry is spontaneously broken down towards the Standard
Model, the broken generators have massive gauge bosons associated with them.
These bosons have masses of the same order as the symmetry breaking scale, ie. 
the vevs of the breaking fields.  Our model already assumes an order for 
symmetry breaking, which creates a vev hierarchy 
$\left( v_{\phi} \geq v_{H} \gg v_{h} \sim O\left( M_{EW} \right) \right)$.  
For instance, we know that $v_{\phi} , v_{H} \gg O\left( M_{EW} \right)$ since 
these broken symmetry bosons have not been observed.

\begin{figure}[h]
 \begin{center}
  \begin{picture}(270,185)(5,5)
   \ZigZag(5,15)(5,35){9}{1} 
   \Line(5,0)(5,15) 
   \LongArrow(5,35)(5,175)
   \Text(5,182)[b]{Energy scale} 
   \Line(2,160)(8,160)          \Text(0,160)[r]{$M \sim m_{s}$} 
   \Text(0,120)[r]{$R_{5_{1}}^{-1}$}
   \Text(0,80)[r]{$R_{5_{2}}^{-1}$} 
   \DashLine(2,120)(270,120){5}
   \DashLine(2,80)(270,80){5}
   \Text(70,140)[]{{\small $d_{5_{1}} = d_{5_{2}} = 6$}} 
   \Text(70,100)[]{{\small $d_{5_{1}} = 4$ , $d_{5_{2}} = 6$}} 
   \Text(70,60)[]{{\small $d_{5_{1}} = d_{5_{2}} = 4$}} 
   \Text(220,140)[]{{\small $g_{5_{1}}, g_{5_{2}}(6d) 
       \sim g_{5_{1}}\left( 2\pi R_{5_{1}} \right), 
        g_{5_{2}} \left( 2\pi R_{5_{2}} \right)$}}
   \Text(220,100)[]{{\small $g_{5_{1}} , g_{5_{2}}(6d) 
    \sim g_{5_{2}} \left( 2\pi R_{5_{2}} \right)$}} 
   \Text(220,60)[]{{\small $g_{5_{1}} , g_{5_{2}}$}}
  \end{picture}
 \end{center}
 \caption{{\small  At energy scales below an inverse compactification 
  radii, the dimension appears too small to observe. The coupling in
  a higher-dimension is related to the same coupling in a lower 
  dimension via eq. (\ref{eq:gscaling}).}}
 \label{fig:couplingscales}
\end{figure}
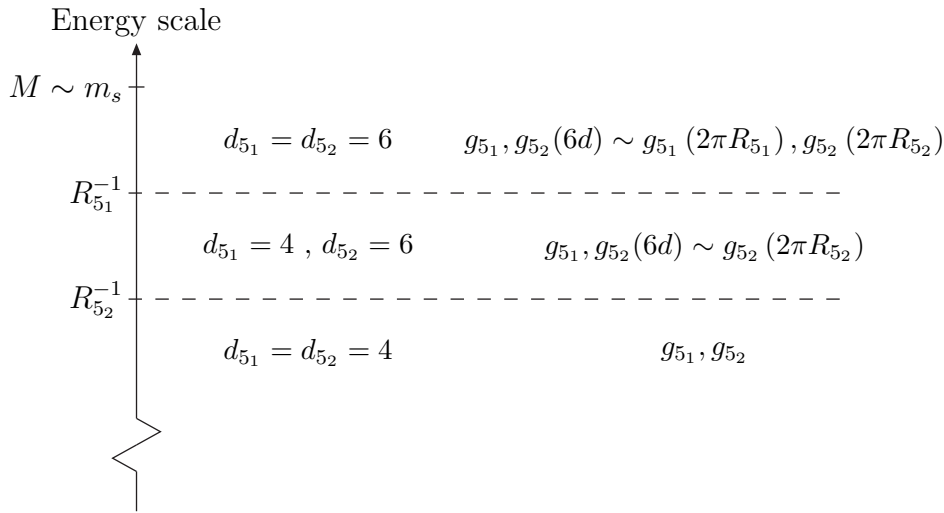

We must also consider the (inverse) compactification radii of the D5-branes.
Their relative sizes are arbitrary, but we choose to start with the
relationship $R_{5_{2}} > R_{5_{1}}$ or equivalently $R_{5_{1}}^{-1} >
R_{5_{2}}^{-1}$ as shown in Fig. \ref{fig:couplingscales}.

Notice that we have not specified how $R_{5_{3}}$ is related
to the other two compactification radii, suffice to say that a large
third dimension (felt by gravity alone) is not forbidden ie.$
R_{5_{1}}^{-1} > R_{5_{2}}^{-1} \gg R_{5_{3}}^{-1} $  (See
\cite{largeexdim} for discussion of large extra dimensions).

In this work, we have adopted the standard scenario with symmetry breaking
occurring at a scale comparable to the first two compactification radii and
string scale.  Soft masses are also generated at around the same scale.
We have deliberately not specified these scales, but we claim
that the formalism applies for GUT/string scales of $1TeV$ to
$10^{16}GeV$.
 
We impose the following restrictions:
\begin{eqnarray}  \label{eq:hierarchy}
    & R_{5_{1}}^{-1} & > R_{5_{2}}^{-1} \nonumber \\
    & v_{\phi} & \geq v_{H} \gg v_{h} \\ 
    & m_{s} & \geq R_{5_{1}}^{-1} , R_{5_{2}}^{-1} \sim v_{\phi} , v_{H} 
     \gg v_{h} \sim O \left( M_{EW} \right) \nonumber
\end{eqnarray}
These constraints provide six ways of ordering the inverse radii and
vevs.  The supersymmetry breaking scale (where soft masses are
generated) also needs to be assigned, thus giving a total of 30
possibilities.
\begin{table}[h]
 \begin{center}
  \begin{tabular}{|c||c|c|c|c|c|c|} \hline
     \vrule width 0pt height 13pt
   A & $v_{h}$ & $v_{H}$ & $v_{\phi}$ & $R_{5_{2}}^{-1}$ & $R_{5_{1}}^{-1}$ 
    & $m_{s}$ \\ \hline
     \vrule width 0pt height 13pt
   B & $v_{h}$ & $v_{H}$ & $R_{5_{2}}^{-1}$ & $v_{\phi}$ & $R_{5_{1}}^{-1}$ 
    & $m_{s}$ \\ \hline
     \vrule width 0pt height 13pt
   C & $v_{h}$ & $R_{5_{2}}^{-1}$ & $v_{H}$ & $v_{\phi}$ & $R_{5_{1}}^{-1}$ 
    & $m_{s}$ \\ \hline
     \vrule width 0pt height 13pt
   D & $v_{h}$ & $v_{H}$ & $R_{5_{2}}^{-1}$ & $R_{5_{1}}^{-1}$ & $v_{\phi}$
    & $m_{s}$ \\ \hline
     \vrule width 0pt height 13pt
   E & $v_{h}$ & $R_{5_{2}}^{-1}$ & $v_{H}$ & $R_{5_{1}}^{-1}$ & $v_{\phi}$ 
    & $m_{s}$ \\ \hline
     \vrule width 0pt height 13pt
   F & $v_{h}$ & $R_{5_{2}}^{-1}$ & $R_{5_{1}}^{-1}$ & $v_{H}$ & $v_{\phi}$ 
    & $m_{s}$ \\ \hline 
  \end{tabular} 
   \caption {{\small Possible ordering of symmetry breaking vevs and inverse 
               compactification radii within the constraints of eq. 
               (\ref{eq:hierarchy}).}}  
 \end{center}
\end{table} 
 
In Table 13 we list the various possibilities for the relative ordering of 
mass scales, vevs and inverse
compactification radii within the constraints of
eq.(\ref{eq:hierarchy}).  
However it is important to notice that when the 
inverse compactification radii are less than
the symmetry breaking vevs, the 
Kaluza-Klein modes associated with the extra dimensions can 
contribute to the running of gauge couplings, leading to a power law 
dependence \cite{dienes}.



\begin{thebibliography}{99}
 \bibitem{gauge} M. Dine, W. Fischler, M. Srednicki, {\it Nucl. Phys.}
                  {\bf B189}, 575 (1981); \\
                 S. Dimopoulos, S. Raby, {\it Nucl. Phys.} {\bf B192}, 353
                  (1981); \\
                 L. Alvarez--Gaum\'e, M. Claudson, M.B. Wise,
                  {\it Nucl. Phys.} {\bf B207}, 96 (1982); \\
                 M. Dine and A.E. Nelson, hep-ph/9303230,
                  {\it Phys. Rev.} {\bf D48}, 1277 (1993); \\
                 M. Dine, A.E. Nelson and Y. Shirman, hep-ph/9408384, 
                  {\it Phys. Rev.} {\bf D51}, 1362 (1995); \\
                 M. Dine, A.E. Nelson, Y. Nir and Y. Shirman,
                  hep-ph/9507378, {\it Phys. Rev.} {\bf D53}, 2658 (1996); \\
                 H. Murayama, hep-ph/9705271, {\it Phys. Rev. Lett.}
                  {\bf 79}, 18 (1997); \\
                 S. Dimopoulos, G. Dvali, R. Rattazzi, G.F. Giudice,
                  hep-ph/9705307,  {\it Nucl. Phys.} {\bf B510}, 12 (1998); \\
                 M.A. Luty, hep-ph/9706554, {\it Phys. Lett.} {\bf 414B}, 71
                  (1997); \\
                 For a review, see G.F. Giudice and R. Rattazzi, 
                  hep-ph/9801271, {\it Phys. Rept.} {\bf 322}, 419 (1999); 
                  {\it Phys. Rept.} {\bf 322}, 501 (1999).
 \bibitem{anomaly} L. Randall and R. Sundrum, hep-th/9810155, {\it Nucl. Phys.} 
                   {\bf B557}, 79 (1999); \\
                   G.F. Giudice, M.A. Luty, H. Murayama and R. Rattazzi, 
                   hep-ph/9810442, {\it JHEP} {\bf 9812}, 027 (1998).
 \bibitem{kaplan} D.E. Kaplan, G. Kribs and M. Schmaltz, hep-ph/9911293, 
                  {\it Phys. Rev.} {\bf D62}, 035010 (2000).
 \bibitem{chacko} Z. Chacko, M. Luty, A.E. Nelson and E. Pont\'{o}n, 
                  hep-ph/9911323, {\it JHEP} {\bf 0001}, 003 (2000).
 \bibitem{no-scale} J. Ellis, K. Enqvist and D.V. Nanopoulos, {\it Phys. Lett.}
                    {\bf B147}, 99 (1984); \\
                    J. Ellis, C. Kounnas and D.V. Nanopoulos, {\it Nucl. Phys. }
                    {\bf B247}, 373 (1984).
 \bibitem{witten} P. Horava and E. Witten, hep-th/9510209, {\it Nucl. Phys.}
                  {\bf B460}, 506 (1996); \\
                  E. Witten, hep-th/9602070, {\it Nucl. Phys.} {\bf B471}, 
                  135 (1996); \\
                  P. Horava and E. Witten, hep-th/9603142, {\it Nucl. Phys.} 
                  {\bf B475}, 94 (1996).
 \bibitem{giudicemasiero} G.F. Giudice and A. Masiero, {\it Phys. Lett.} {\bf 206B}
                          , 480 (1988).
 \bibitem{shiutye} G. Shiu and S.-H. Henry Tye, hep-ph/9805157, {\it Phys. Rev.}
                   {\bf D58}, 106007 (1998).
 \bibitem{dienes} K.R. Dienes, E. Dudas and T. Gherghetta, hep-ph/9806292, 
                  {\it Nucl. Phys.} {\bf B537}, 47 (1999);
I. Antoniadis, {\it Phys. Lett.} {\bf B246}, 377 (1990).
 \bibitem{kaku} Z. Kakushadze, hep-th/9806008, {\it Nucl. Phys.} {\bf B535}, 
                311 (1998).
 \bibitem{ibanez} L.E. Ib\'{a}\~{n}ez, C. Mu\~{n}oz and S. Rigolin, hep-ph/9812397,
                  {\it Nucl. Phys.} {\bf B553}, 43 (1999).
\bibitem{gordy}
L.Everett, G.L.Kane and S.F.King, hep-ph/0005204, {\it JHEP} {\bf 008}, 012
 (2000).
 \bibitem{king} S.F. King and M. Oliveira, hep-ph/0009287, {\it Phys. Rev.}
 {\bf D63}, 095004 (2001).
 \bibitem{largeexdim} N. Arkani-Hamed, S. Dimopoulos and G. Dvali,
        hep-ph/9803315, {\it Phys. Lett.} {\bf B429}, 263 (1998);
I. Antoniadis, N. Arkani-Hamed, S. Dimopoulos and G. Dvali,
        hep-ph/9804398, {\it Phys. Lett.} {\bf B436}, 257 (1998).
\bibitem{LEP}
The ALEPH Collaboration, hep-ex/0011045, {\it Phys. Lett.} {\bf B495}, 1
 (2000);
The L3 Collaboration, hep-ex/0011043, {\it Phys. Lett.} {\bf B495}, 18 (2000).
\bibitem{Higgs}
G.L.Kane, S.F.King and L-T. Wang, hep-ph/0010312;
J.Ellis, G.Ganis, D.V.Nanopoulos and K.A.Olive, hep-ph/0009355, {\it Phys. 
 Lett.} {\bf B502}, 171 (2001).
\bibitem{Wang} M. Cvetic, M. Plumacher and J. Wang, hep-th/9911021,
                {\it JHEP} {\bf 0004}, 004 (2000); \\
                M. Cvetic, A.M. Uranga and J. Wang, hep-th/0010091.
\bibitem{Benakli} K. Benakli, hep-ph/9911517, {\it Phys. Lett.}
                  {\bf B475}, 77 (2000).

\end{thebibliography}
\end{document}